\def\Msun{M$_\odot$}

\def\kms    {\ifmmode{{\rm \ts km\ts s}^{-1}}\else{\ts km\ts s$^{-1}$}\fi}

\documentclass{aa}  
\usepackage{graphicx}
\usepackage{lscape}
\usepackage{txfonts}
\usepackage{color}
\begin{document}
   \title{A very extended molecular web around NGC~1275}

   \author{P. Salom\'e
          \inst{1}
          \and
          F. Combes\inst{1}\fnmsep
          \and
          Y. Revaz\inst{5}\fnmsep
          \and
          D. Downes\inst{2}\fnmsep
          \and
          A.C. Edge\inst{3}\fnmsep
          \and
          A.C. Fabian\inst{4}\fnmsep
}

   \institute{LERMA, Observatoire de Paris, UMR 8112 du CNRS, 
             75014 Paris, France
      \email{philippe.salome@obspm.fr}
   \and
        Institut de Radio Astronomie Millim\'etrique, Domaine
        Universitaire, 38406 St.\ Martin d'H\`eres, France
        \and
            Institute of Computational Cosmology, Durham University,
             South Road, Durham, DH1 3LE, UK
         \and
             IoA, Madingley Road, Cambridge, CB3 OHA, UK
         \and             
               Laboratoire d'Astrophysique, \'Ecole Polytechnique 
             F\'ed\'erale de Lausanne (EPFL), Observatoire, 1290 
             Sauverny, Switzerland}

   \date{Received November 12th, 2008; accepted May 16th, 2011}

\abstract{We present the first detection of CO emission lines in the 
H$\alpha$ filaments at distances as far as 50 kpc from the centre of
the galaxy NGC~1275.  This gas is probably dense
($\ge$10$^{3}$cm$^{-3}$). However, it is not possible to accurately
determine {the density and the} kinetic temperature of this relatively
warm gas (T$_{\rm kin}$$\sim$\,20-500\,K) with the current data
only. The amount of molecular gas in the filaments is large ---
10$^9$\,\Msun (assuming a Galactic N(H$_2$)/I$_{\rm CO}$ ratio).  This
is 10\% of the total mass of molecular gas detected in this cD galaxy.
This gas has large-scale velocities comparable to those seen in
H$\alpha$.  The origin of the filaments is still unclear, but their
formation is very likely linked to the AGN positive feedback (Revaz
et al., 2008) that regulates the cooling of the surrounding
X-ray-emitting gas as suggested by numerical simulations. We also
present high-resolution spectra of the galaxy core. The spatial
characteristics of the double-peaked profile suggest that the
molecular web of filaments and streamers penetrates down to radii of
less than 2 kpc from the central AGN and eventually feed the galaxy
nucleus. The mass of gas inside the very central region is
$\sim$10$^9$\,\Msun, and is similar to the mass of molecular gas found
in the filaments.

\keywords{galaxies: cooling flows  --- galaxies: individual: NGC~1275
      --- galaxies: ISM --- galaxies: kinematics and dynamics}
}
   \maketitle
%
%---------------------------------------------------------------------------
\section{Introduction}
%---------------------------------------------------------------------------

The cD galaxy NGC~1275 lies at the centre of the Perseus cluster, the
brightest galaxy cluster in the sky in X-rays. The galaxy is at a
redshift of 0.01756, ($D_A$ = 72.6\,Mpc; 1$''$ is 350 pc).  Past
detections of CO in NGC~1275 were mainly toward the centre of the
galaxy (Lazareff et al., 1989; Mirabel et al, 1989; Reuter et al.,
1993; Braine et al., 1995; Inoue et al, 1996; Bridges \& Irwin, 1998;
Lim et al., 2008).  We also studied CO in filamentary structures all
around the galaxy (inside r$\le$10 kpc) and in three pointings at
galactocentric distances as large as r$\sim$25 kpc (Salom\'e et al.,
2006, 2008a, 2008b).  The molecular gas coincides with the H$\alpha$
filaments (e.g., Cowie et al. 1983; Conselice et al. 2001), within the
hot gas seen in soft X-rays at 0.5 keV (Fabian et al. 2006). {These
results were in good agreement with the CO observations of the central
15 kpc made by Reuter et al., (1993) although the regions observed
were not exactly the same. The central region was also observed with
the VLA by Jaffe (1990) who detected extended 21cm absorption in front
of the radio source 3C84. The HI is elongated in a east-west direction
like the CO and the inferred neutral hydrogen mass found is
$\sim$5\,10$^{9}$M$_\odot$, that is, of the same order as the
molecular gas mass ($\sim$10$^{10}$M$_\odot$) found in the same region
from CO observations (Salom\'e et al. 2006). {Irwin \& Bridges (2001)
found a north-west/south-east extended continuum emission at 450$\mu$m
and 850$\mu$m with SCUBA. This core-subtracted emission excess reveals
a large amount of cold dust inside the central $\sim$20 kpc
(6$\times$10$^7$M$_\odot$ with T$_{\rm dust}$$\sim$20K and
$\beta$=1.3)}.

In this paper, we describe a new search with the IRAM 30m telescope
for molecular gas in the H$\alpha$ filaments at the largest radii from
the centre of NGC~1275 (outside the already observed central region
and as far as 50 kpc). We clearly identify molecular filaments in
five regions (see Fig.~1):

\noindent
(1) A 30\,kpc-long filament, running north-south, not previously
observed in CO (regions r1 through r4).  The largest radius at which
we detect CO (region r1) is 50\,kpc (in projection) north of the
centre of the galaxy.

\noindent
(2) The southeast H$\alpha$ filaments (r10, r11, r15). The distance
between the end of this filamentary system (region r11) and the end of
the north filament (r1) is 200$''$ (70\,kpc),

\noindent
(3) A tangential filament to the north-east (r19, r20), where we have
extended our previous CO observations.

\noindent
(4) The northwest area, including the well-known horseshoe filament,
was observed in regions r5, r6, r8, r9, and r16.

\noindent
(5) The central region around 3C84 was re-observed.  Here, we made a
high-sensitivity spectrum with high spectral resolution.

Section 2 presents the observations. Section 3 describes the new CO
detections. Section 4 discusses the excitation conditions of the
molecular gas at large radii around NGC~1275 and the possible origin
of the very distant molecular filaments.\\

\vspace{-0.55cm}

\section{Observations}
We used the IRAM 30\,m telescope on Pico Veleta, near Granada,
Spain. Observations were made in December 2007. Receivers were tuned
to the CO(1--0) and CO(2--1) lines, redshifted to the adopted systemic
velocity of NGC~1275 (113.280 and 226.559\, GHz). We used two
512\,$\times\,$1\,MHz filters at 3\,mm and two 250\,$\times\,$\,4 MHz
filters for the 1.3\,mm receivers, thereby covering 1300\,km/s at both
wavelengths.  We used wobbler switching, and checked the pointing on
3C84, the radio source at the centre of NGC~1275.  We observed in
excellent weather with T$_{{\rm sys}}$ typically 250\,K at 1.3\,mm.
To study the large extent of molecular gas around NGC~1275, we
observed 18 regions inside the H$\alpha$ filaments
(Fig \ref{overlay}). The beams of the 30\,m telescope, at 3\,mm and
1.3\.mm, are 22$^{\prime\prime}$ and 11$^{\prime\prime}$ respectively
(corresponding to 7.7 and 3.8\,kpc at NGC~1275).  The line intensities
are given in main-beam brightness temperatures, $T_{mb}$.  The ratio
of $T_{mb}$ to $T_a^*$ at the 30\,m is $T_{mb}/T_a^* = F_{\rm
eff}/B_{\rm eff}$ = 1.27 at 3\,mm and 1.75 at 1.3\,mm (see
http://www.iram.es/).

The data were calibrated with the MIRA software and reduced with the
CLASS package. We dropped spiky channels and bad scans, and subtracted
linear baselines from each spectrum before averaging.
Table \ref{table-results} summarises the results.

%%%%%%%%%%%%%%%%%%%%%%%% CO(1-0) and CO(2-1) %%%%%%%%%%%%%%%%%%%%%%%%%%%%%%%
%\begin{landscape} 
\begin{figure*} \centering
\begin{tabular}{c}
\includegraphics[width=19cm, angle=-0]{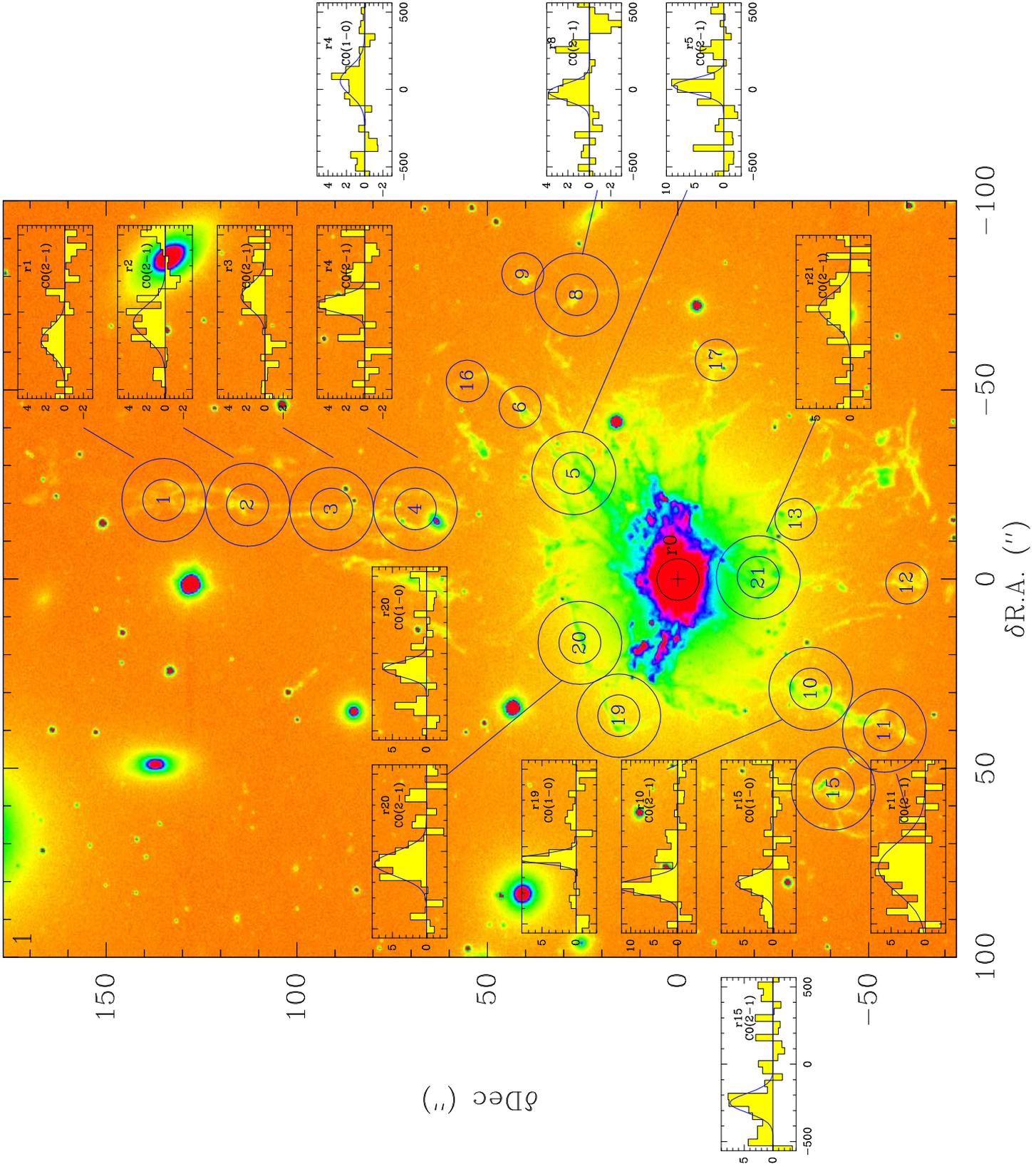}\\
\end{tabular}
\caption{CO spectra superposed on the 
H$\alpha$ filament structure around NGC~1275 (Conselice et al., 2001).
The double circles show the 1.3 and 3\,mm beamwidths at the regions
where CO was detected. CO(1--0) and CO(2--1) spectra are shown when
detected. The channel width is 42 km/s and the y-axis is main beam
brightness temperature, in mK. The velocity scale is from $-$560 to
560\,km/s in each insert. Small circles (1.3\,mm beamwidth) show
regions where no CO was detected. See Table
\ref{table-results} for more details.}
\label{overlay} 
\end{figure*} 
%\end{landscape} 
%%%%%%%%%%%%%%%%%%%%%%%%%%%%%%%%%%%%%%%%%%%%%%%%%%%%%%%%%%%%%%%%%%%%%%%%%%%%

%---------------------------------------------------------------------------

%%%%%%%%%%%%%%%%%%%%%%%% RESULTS SUMMARY TABLE  %%%%%%%%%%%%%%%%%%%%%%%%%%%%%
%
%
\begin{table*}[!htbp]
\begin{center}
\caption{CO detections and tentative detections, or 3$\sigma$ upper limits. See Salom\'e et al. (2008a) 
for complementary data.}
\begin{tabular}{cc ccc ccc} 
\hline 
\hline 
Site & Offsets &CO &$T_{mb}$ &$V$  &$\Delta$V &$I_{\rm CO}$ &$M_{\rm gas}$ \\ 
    & $''\times''$ &line &mK           &km/s &km/s &K\,km/s &10$^8$\,M$_\odot$  \\ 
\\ 
\hline 
\multicolumn{7}{l}{\itshape Northern Filament} \\
\hline 
r1 &-21.0$\times$134.9& CO(2--1) & 2.5 $\pm$ 1.2 &-184.7 $\pm$ 28.2 &198.9 $\pm$ 52.9 &0.5 $\pm$ 0.1 & 0.3 \\ 
r1 && CO(1--0) &  $<$ 3$\times$0.6 & - & - & - & -  \\ 
\hline 
r2 &-19.7$\times$112.9& CO(2--1) & 3.5 $\pm$ 1.2 &-65.3 $\pm$ 23.4 &246.4 $\pm$ 43.4 &0.9 $\pm$ 0.2 & 0.5 \\ 
r2 && CO(1--0) &  $<$ 3$\times$1 & - & - & - & -  \\ 
\hline 
r3 &-18.6$\times$90.9& CO(2--1) & 2.6 $\pm$ 1.2 &100.2 $\pm$ 24 &187.6 $\pm$ 48 &0.5 $\pm$ 0.1 & 0.3 \\ 
r3 && CO(1--0) &  $<$ 3$\times$1 & - & - & - & -  \\ 
\hline 
r4 &-18.6$\times$68.9& CO(2--1) & 5.7 $\pm$ 1.5 &51.3 $\pm$ 12.6 &112.9 $\pm$ 23.3 &0.7 $\pm$ 0.1 & 0.4 \\ 
r4 && CO(1--0) & 2.7 $\pm$ 0.8 &55.2 $\pm$ 19.4 &209.3 $\pm$ 38.8 &0.6 $\pm$ 0.1 & 1.4 \\ 
\\
\hline 
\multicolumn{7}{l}{\itshape South East Filament}\\
\hline 
r10 &28.8$\times$-34.9& CO(2--1) & 13.6 $\pm$ 2 &-255.8 $\pm$ 8 &114.5 $\pm$ 23.2 &1.7 $\pm$ 0.2 & 1 \\ 
r10 && CO(1--0) &  $<$ 3$\times$2 & - & - & - & -  \\ 
\hline 
r11 &39.8$\times$-54.2& CO(2--1) & 6.9 $\pm$ 1.7 &-137.5 $\pm$ 19.4 &373.1 $\pm$ 47.6 &2.7 $\pm$ 0.3 & 1.6 \\ 
r11 && CO(1--0) &  $<$ 3$\times$1.8 & - & - & - & -  \\ 
\hline 
r15 &55.2$\times$-40.8& CO(2--1) & 7.5 $\pm$ 2.4 &-250.6 $\pm$ 21.6 &146.8 $\pm$ 67.2 &1.2 $\pm$ 0.4 & 0.7 \\ 
r15 && CO(1--0) & 6.6 $\pm$ 1.7 &-241.4 $\pm$ 15.5 &123.5 $\pm$ 30.7 &0.9 $\pm$ 0.2 & 2 \\ 
\\
\hline 
\multicolumn{7}{l}{\itshape North East Tangential Filament}\\
\hline 
r19 &35.9$\times$15.5& CO(2--1) &  $<$ 3$\times$1.7 & - & - & - & - \\ 
r19 && CO(1--0) & 8.9 $\pm$ 1.3 &-77.7 $\pm$ 5.4 &68.8 $\pm$ 10.7 &0.7 $\pm$ 0.1 & 1.5 \\ 
\hline 
r20 &16.7$\times$25.7& CO(2--1) & 7.7 $\pm$ 1.7 &-79.8 $\pm$ 14.6 &184.7 $\pm$ 28.6 &1.5 $\pm$ 0.2 & 0.9 \\ 
r20 && CO(1--0) & 6.4 $\pm$ 1.6 &-100.4 $\pm$ 12.2 &109.4 $\pm$ 24.3 &0.7 $\pm$ 0.2 & 1.7 \\ 
\\
\hline 
\multicolumn{7}{l}{\itshape Horseshoe Filament}\\
\hline 
 r5 &-28.1$\times$27.3& CO(2--1) & 9.1 $\pm$ 2.3 &23.5 $\pm$ 13.6 &101.5 $\pm$ 58.0 &1.0 $\pm$ 0.3 & 0.6 \\ 
r5 && CO(1--0) &  $<$ 3$\times$1.6 & - & - & - & -  \\ 
\hline 
r6 &-45.6$\times$41.4& CO(2--1) &  $<$ 3$\times$2.7 & - & - & - & - \\ 
r6 && CO(1--0) &  $<$ 3$\times$2 & - & - & - & -  \\ 
\hline 
r8 &-75.3 26.5& CO(2--1) & 3.9 $\pm$ 1.5 &-17.1 $\pm$ 18.9 &122.1 $\pm$ 35.2 &0.5 $\pm$ 0.1 & 0.3 \\ 
r8 && CO(1--0) &  $<$ 3$\times$0.9 & - & - & - & -  \\ 
\hline 
r9 &-80.8$\times$40.7& CO(2--1) &  $<$ 3$\times$1.6 & - & - & - & - \\ 
r9 && CO(1--0) &  $<$ 3$\times$0.9 & - & - & - & -  \\ 
\hline  
r16 &-52.5$\times$55.2& CO(2--1) &  $<$ 3$\times$1.5 & - & - & - & - \\ 
r16 && CO(1--0) &  $<$ 3$\times$1.2 & - & - & - & -  \\ 
\\
\hline 
\multicolumn{7}{l}{\itshape Southern Filament}\\
\hline 
r12 &0.8$\times$-60.1& CO(2--1) &  $<$ 3$\times$1.7 & - & - & - & - \\ 
r12 && CO(1--0) &  $<$ 3$\times$1.6 & - & - & - & -  \\ 
\hline 
r13 &-15.9$\times$-30.9& CO(2--1) &  $<$ 3$\times$1.8 & - & - & - & - \\ 
r13 && CO(1--0) &  $<$ 3$\times$1.3 & - & - & - & -  \\ 
\hline 
r21 &-0.6$\times$-21.1& CO(2--1) & 4.7 $\pm$ 2.3 &86.7 $\pm$ 37.2 &222 $\pm$ 78.3 &1.1 $\pm$ 0.4 & 0.6 \\ 
r21 && CO(1--0) &  $<$ 3$\times$3.2 & - & - & - & -  \\ 
\\
\hline 
\multicolumn{7}{l}{\itshape Western Filament}\\
\hline 
r17 &-58.0$\times$-10.1& CO(2--1) &  $<$ 3$\times$1.8 & - & - & - & - \\ 
r17 && CO(1--0) &  $<$ 3$\times$1 & - & - & - & -  \\ 
\\
\hline 
\multicolumn{7}{l}{\itshape Center} \\ 
\hline 
r0 & 0$\times$0 & CO(2--1) &  72 $\pm$ 6.9 & 4  $\pm$ 4 & 293 $\pm$ 10 & 22.5 $\pm$ 0.6 &  15.3 \\ 
\end{tabular}
\label{table-results}
\end{center}
T$_{mb}$ = main-beam temperature at line peak; 
$V$ = velocity relative to systemic; 
$\Delta$V = linewidth (FWHM); 
$I_{\rm CO}$ = integrated line intensity; 
$M_{\rm gas}$(H+He) is for $M/L^\prime_{\rm CO}$ =
4.6\,\Msun\,(K\,\kms\,pc$^2$)$^{-1}$ (Solomon et al., 1997). Note that the integration time
is not the same for all regions, so the rms (computed for a spectral 
resolution of 42 km/s) varies from one pointing to another. 
\end{table*}
 
%%%%%%%%%%%%%%%%%%%%%%%%%%%%%%%%%%%%%%%%%%%%%%%%%%%%%%%%%%%%%%%%%%%%%%%%%%%%
%%---------------------------------------------------------------------------
\section{Results, by region} 
%---------------------------------------------------------------------------
\subsection{The gas masses}

For comparison with other work, the gas mass was estimated from a
standard Milky Way conversion factor M$_{\rm gas}$/L$^\prime_{\rm
CO}$\,=\,4.6\, M$_\odot$\,(K.km/s.pc$^{2}$)$^{-1}$ (Solomon et
al. 1997) for both transitions. Note however that the ICM metallicity
is known to be less than solar Z$<$0.6Z$_\odot$ in the Perseus cluster
core (Schmidt et al., 2002). This low metallicity would lead to a
larger N(H$_2$)/CO conversion factor than the standard one used here
(Leroy et al, 2009) and thus under-estimated gas masses. The CO
luminosity is defined by L$^\prime_{\rm CO}$\,= T$_{\rm
b}$\,$\Delta$V\,$\Omega_S$\,D$^2_A$ with T$_{\rm b}$\,$\Omega_S$ the
source brightness temperature times the source solid angle, $\Delta$V
the line width and D$^2_A$ the angular distance of the source. This
gives L$^\prime_{\rm CO}$\,=\,23.5\,$\Omega_{\rm s*b}$\,I$_{\rm
CO}$\,D$_{\rm A}^2$\,(1+z), with z the source redshift, $\Omega_{\rm
s*b}$ the solid angle of the source convolved with the telescope beam
and I$_{\rm CO}$=$\int$T$_{\rm mb}\,\Delta$V the CO intensity (T$_{\rm
mb}$ being the main beam temperature). This is for L$^\prime_{\rm CO}$
in K.km/s.pc$^2$, $\Omega_{\rm s*b}$ in arcsec$^2$, D$_{\rm A}$ in
Mpc, and I$_{\rm CO}$ in K.km/s.  With this conversion factor, we find
molecular gas masses of $\sim$3-16$\times$10$^7$\,M$_\odot$ in each
pointed region of the filaments. We used the same conversion factor
for both transitions since line ratios were close to one. We find more
mass in the (larger) 3mm beam than in the 1mm beam. So the CO emission
may be extended. Note however that this is just a hint because of the
large uncertainties on the line width and main beam temperature
measurements for both transitions.\\

\subsection{The 30-kpc-long north filament, at very large radius:} This
filament extends over a radial distance from 20 to 50\,kpc from the
galaxy's centre. We detected CO(2--1) at all positions observed, but
no CO(1--0) apart from r4 (Table \ref{table-results}, regions r1
to r4).  The fact that CO(2--1) is easier to detect means that the
CO(1--0) is strongly affected by beam dilution, and therefore the
emission is fragmented and clumpy on a scale of a few kpc (in region
r4, where both lines are detected, the CO(2--1)/CO(1--1) main beam
temperature ratio is $\sim$2).  The CO(2--1) line centre changes from
$-$184\,km/s (at r1), to $-$65\,km/s (at r2), then jumps to +100\,km/s
(r3), and +50\,km/s (r4).  These line velocities show the same trend
as in H$\alpha$ (Hatch et al. 2006). Figure \ref{simu} (middle panel)
shows the observed position-velocity diagram along this filament.
There is a positive gradient from $-$200\,km/s to +100\,km/s as the
radius decreases from 50 to 35\,kpc, after which the velocity falls
back to 40\,km/s at a radius of 25\,kpc.  The good agreement with the
H$\alpha$ suggests that the CO traces most of the mass in the
filaments, possibly surrounded by ionised gas radiating in H$\alpha$,
at the interface with the hot intra-cluster medium.  The shape of the
gradient and the velocity reversal at the end could be due to the
projection of the curved filament.  Judging from the other loops that
are oriented more face-on, it is obvious that many of the filaments
have a significant curvature.  The linewidth of regions r1 to r3 is
200\,km/s while region r4 shows a narrower linewidth of 100\,km/s,
more typical of the sub-structures in the eastern filament (see
Salom\'e et al. 2008b). With the standard Milky Way conversion factor
described above, we find a molecular gas mass of
$\sim$3-14$\times$10$^7$\,M$_\odot$ in the northern filament.

\subsection{The bright southeast filaments:} Regions r10, r11 and r15
(the 'blue-loop') cover the southeast H$\alpha$ filaments. Regions r10
and r11 are only detected in CO(2--1), while region r15 is also
detected in CO(1--0). This filament has negative velocities
only. There is a velocity change from $-$130\,km/s to $-$250\,km/s
between regions r11 and r10 (compatible with the velocities of a
falling filament). Region r11 has a broad CO(2--1) linewidth
(370\,km/s), while in region r10, the CO(2--1) profile is narrow
(120\,km/s).  Slightly farther east is the region r15 at a velocity of
$-$\,130 km/s. These southeast H$\alpha$ filaments are wider than the
Northern ones, and the CO signal is probably less affected by beam
dilution.

%---------------------------------------------------------------------------
\subsection{The north-east tangential filament:}
This filament is not radially extended like most of the other
filaments around NGC~1275, but is instead perpendicular to a line
through the centre of the galaxy. Region r19 is detected in CO(1--0)
and region r20 in both CO(1--0) and CO(2--1). The line velocities are
all negative, with no clear gradient. The line widths are narrower
(70, 100 and 180\,km/s) than those of the other detections.

%---------------------------------------------------------------------------
\subsection{The horseshoe: the faintest region:}
The horseshoe was re-observed in different regions (r5, r6, r8, r9,
r16) all along both sides of the U-shaped loop. We clearly detect the
CO(2--1) line in the trunk of the filament, but this is the only firm
detection.  CO(2--1) is tentatively detected in region r8.  The other
regions were not detected in CO(1--0) or CO(2--1). Surprisingly, this
region has weaker CO emission than the other filaments. If the
inclination of the horseshoe filament along the line of sight is
large, then the regions we observe could be at galactocentric
distances as far as the faint regions detected in the Northern
filament (50 kpc). In such a case, the small amount of molecular
gas would not be visible with the rms noise we reached for those
regions. The only exception is in region r8 where the integration time
was longer (see Table 1).

%---------------------------------------------------------------------------

%%%%%%%%%%%%%%%%%%%%%%%%%%%%%%%%%%%%%%%%%%%%%%%%%%%%
% LVG 
%%%%%%%%%%%%%%%%%%%%%%%%%%%%%%%%%%%%%%%%%%%%%%%%%%%%
\begin{figure*}[htbp]
\centering    
\includegraphics[width=6.5cm,angle=-90]{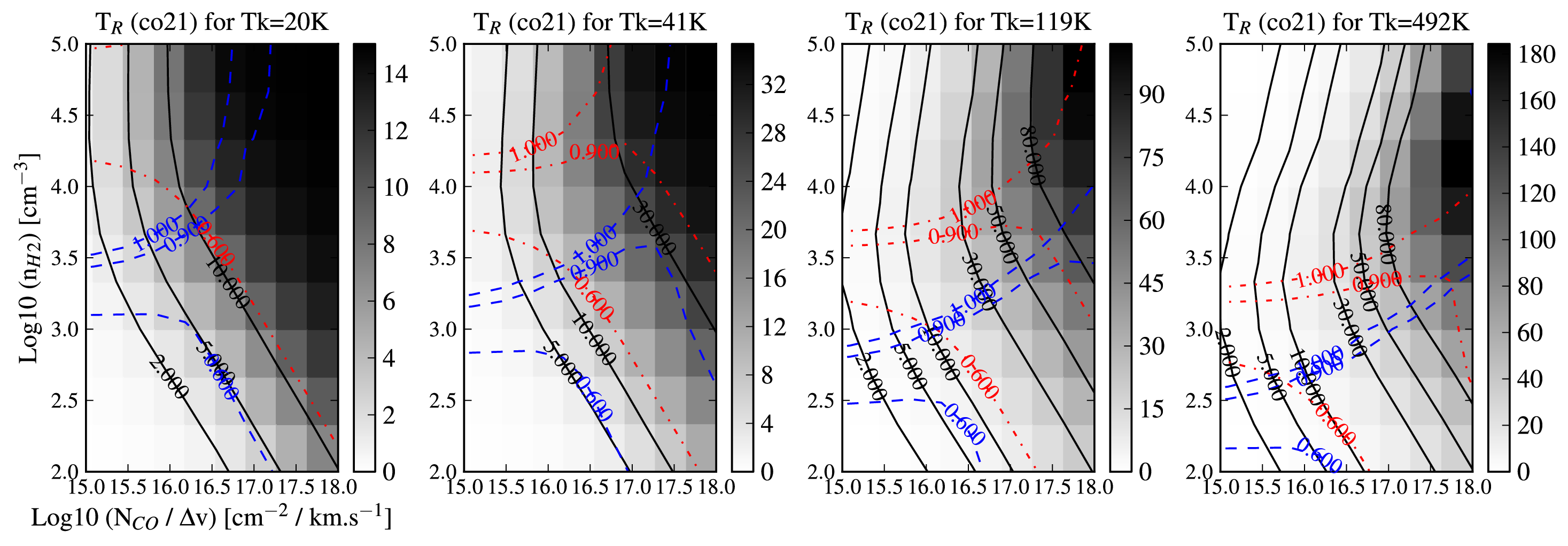}
\caption{Results of escape probability models from the 
RADEX program (Van der Tak et al., 2006) for {four} different kinetic
 temperatures, a range of H$_2$ densities and a range of CO molecular
 column density per km/s. The grey scale mapping and the black contour
 lines show the expected CO(2--1) emission T$_R$(co$_{21}$) [K]
 (see wedge to the right of each panel).  Dashed blue lines show the
 CO(2--1)/CO(1--0) line temperature ratios 0.6, 0.9 and 1 as predicted
 by the RADEX program.  Dash-dot red lines show the predicted
 CO(3--2)/CO(2--1) line temperature ratios 0.6, 0.9 and 1. The
 measured CO(2--1)/CO(1--0) line temperature ratio in r15 and r20 are
 close to one. So the possible solutions lie along the dashed-blue
 line of value one. We see that it is thus difficult to constrain the
 range of possible kinetic temperatures and H$_{\rm 2}$
 densities. Nervertheless, even with a high T$_{\rm kin}$ of 500\,K,
 the density must be $\ge$10$^{2.5}$. Note also that if we consider a
 standard Milky Way M$_{\rm gas}$/L$'_{\rm co}$ conversion factor,
 then we expect N$_{\rm
 CO}$/$\Delta$V$>$1\,10$^{17}$\,cm$^{-2}$km.s$^{-1}$ if f$_{\rm
 2D}\sim$10$^{-3}$ ie T$_{\rm R}\sim$10\,K, which gives slightly
 higher densities ($\ge$10$^{2.9}$).  Finally, Bridges \& Irwin (1998)
 reported a CO(3--2)/CO(2--1) line ratio close to one in the central
 $\sim$ 8kpc region. If this is also true for the filaments, then the
 possible solutions would lie at the intersection of the dashed blue
 line (values $\sim$0.9-1) with the dash-dot red line (values
 $\sim$0.9-1), which means even higher densities. The high-J CO lines
 are thus important diagnostics to determine the gas properties.}
\label{lvg}
\end{figure*}

%---------------------------------------------------------------------------

\subsection{Gas in the central 4\,kpc:}
Figure \ref{center-co21} shows a double-peaked CO(2--1) line at the
position of 3C84 in 5\,\kms\ channels in a 4-kpc diameter region. The
negative velocities peak at -45\,\kms\ and the positive ones at
+65\,\kms\. This feature can be interpreted as (i) inflowing gas, (ii)
outflowing gas or (iii) a rotating pattern as observed in lines in the
NIR by Wilman et al. (2005).  It is not possible to disentangle from
this three possibilities with the present observations only. However,
in previous observations (Salom\'e et al., 2006), the absence of a
larger-scale rotation pattern was pointed out (inside a 8-kpc
diameter region).
%%%%%%%%%%%%%%%%%%%%% ROTATION CENTRAL PIXEL CO21  %%%%%%%%%%%%%%%%%%%%%%%%%%
\begin{figure}
\begin{center}
\includegraphics[width=4cm, angle=-90]{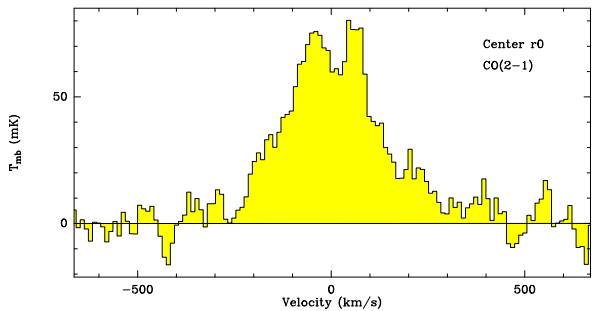}
\caption{CO(2--1) at the position of 3C84. The axes are 
main-beam brightness 
temperature (in mK) vs.\ velocity, relative to 226.559\,GHz
($z$ = 0.01756).  
The channel width is 5\,\kms .
}
\label{center-co21}
\end{center}
\end{figure}
%%%%%%%%%%%%%%%%%%%%%%%%%%%%%%%%%%%%%%%%%%%%%%%%%%%%%%%%%%%%%%%%%%%%%%%%%%%%%

The double-peaked line cannot therefore be unambiguously interpreted
as evidence for a central rotating disk.  Instead, the large
accumulation of molecular gas (nearly 10$^{9}$\,\Msun) in the central
4-kpc diameter of the galaxy could be the result of the recycling of
cold gas inflow from the filaments back into the central potential
well over several Gyr.

%
%---------------------------------------------------------------------------
\section{Discussion}
%---------------------------------------------------------------------------
%%%%%%%%%%%%%%%%%%%%% ROTATION CENTRAL PIXEL CO21  %%%%%%%%%%%%%%%%%%%%%%%%%%
\begin{figure*}[htbp]
\begin{center}
\includegraphics[width=18cm]{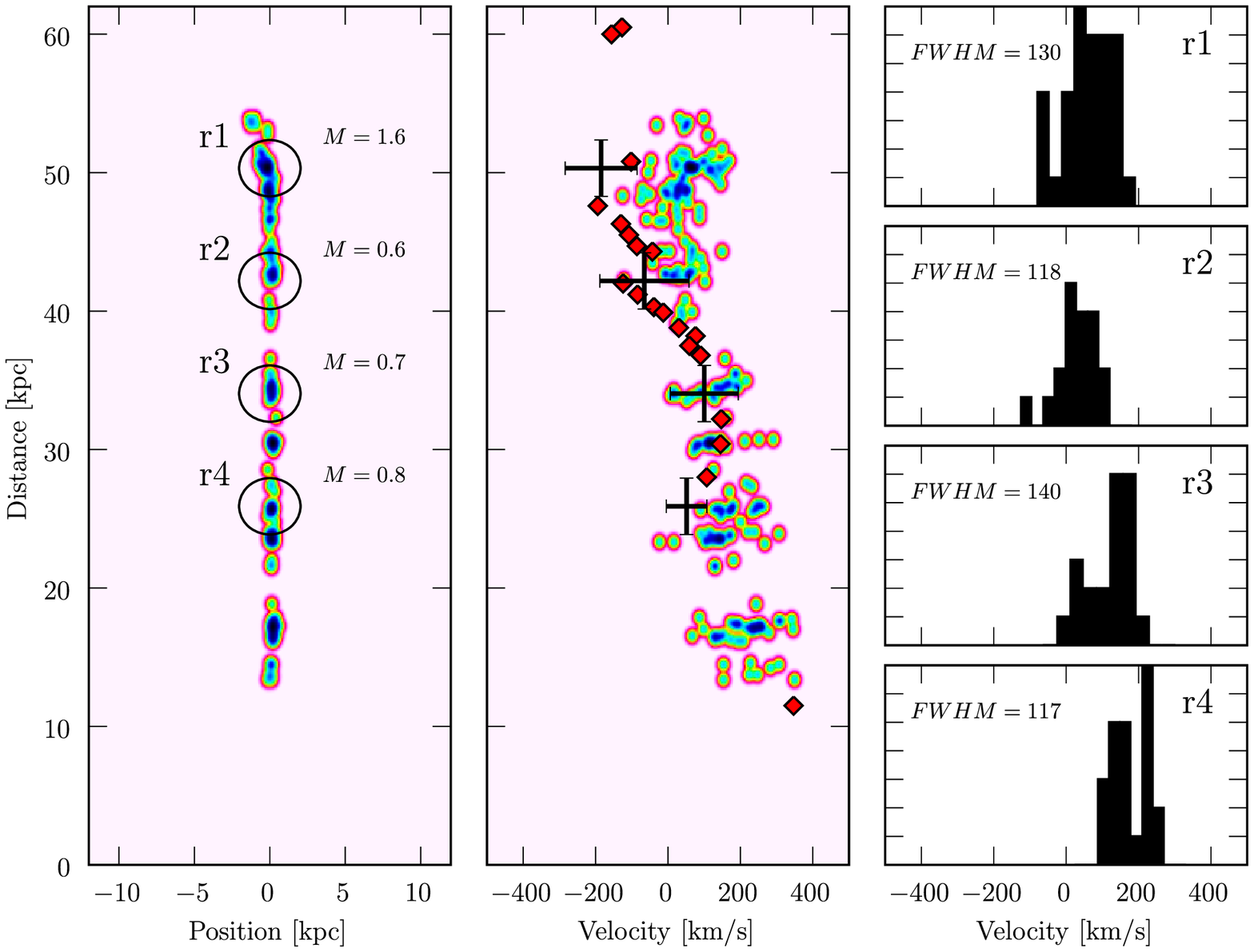}
\caption{Northern filament: comparison of the model with the observations.
{\it Left:} contours are the surface density of molecular gas,
predicted from the model.  CO(2--1) beamwidths and positions are shown
by black circles.  Gas masses in each beam, predicted from the model,
are in units of $10^8\,\rm{M_\odot}$.
{\it Middle:} position-velocity diagram.  Pink and blue isophotes in
the color map are the model predictions.  Black crosses are CO(2--1)
observations.  Horizontal bars are the FWHM linewidths.  Red diamonds
are H$\alpha$ data (Hatch et al. {2006}).
{\it Right:} velocity distribution of the gas extracted from each
region of the model.  The predicted FWHM linewidths of the
distributions are in km/s.  }
\label{simu}
\end{center}
\end{figure*}
 
\subsection{Large velocity dispersions}
Observations by Salom\'e et al. (2008a) showed CO linewidths of 10 to
30\,\kms\ in the eastern filament when observed with a 3$''$ beam.  In
our 22$''$ beam, however, the CO(1--0) linewidth in that same filament
is close to 100\,\kms . In the current data from the 30\,m telescope,
most of the CO(1--0) and (2--1) lines have widths of order 200\,\kms\
in beams of 4 to 8\,kpc. In the northern filament, the velocity
dispersion is even larger. These large linewidths are difficult to
explain by motions within a single filament. It is more likely that in
our beam, we observe an ensemble of very thin sub-filaments (as
observed in the optical with the HST by Fabian et al. 2008) that
consist of clumps of molecular clouds. Furthermore, these individual
sub-filaments are very likely to have different projection angles,
which would also increase the observed linewidths, when observed with
a larger beam. Nevertheless, region r11 has a particularly broad line
width (373 km/s). More observations with a larger bandwith coverage,
as provided now by EMIR (Eight Mixer
receiver)\footnote{http://www.iram.es/IRAMES/mainWiki/EmirforAstronomers}
on the IRAM 30m-telescope, would be useful to constrain with more
confidence the effect of baseline fluctuation to which the results are
very sensitive in the low signal/noise regime.

%%%%%%%%%%%%%%%%%%%%%%%%%%%%%%%%%%%%%%%%%%%%%%%%%%%%%%%%%%%%%%%%%%%%%%%%%%%%%

\subsection{Comparison with numerical models} 
In Figure \ref{simu}, we compare the properties of the CO in the
northern filament with the prediction of models in which the molecular
gas forms out of uplifted ambient gas trapped by a plasma bubble.  We
have modelled the AGN feedback by instantaneously generating bubbles
of gas in the ICM as already done by several authors (see Br\"uggen \&
Kaiser 2001, for instance). We have used very-high resolution TreeSPH
simulations combined with a multiphase model and a model of plasma
bubbles.  We have also taken into account the presence of cooler and
denser around the radio bubbles, assuming that it is the result of
ambient gas pushed out by the inflating bubble.  In our simulations, a
fraction of the 1$-$2 keV gas present at the centre of clusters is
trapped and entrained by the rising buoyant bubble to higher radius
where the AGN heating is less efficient. The radiative cooling then
makes it cool, thereby forming cold filamentary structures in the wake
and in the rim of the bubbles (see Revaz et al.\, 2008 for a detailed
description of the model).  We tried to fit the data presented here
by the model number 2 of Revaz et al.\ (2008). This is at
$t=260\,\rm{Myr}$ and for a filament that is inclined by $60^{\circ}$
with respect to the line of sight (the top of the filament pointing
towards the observer). The model reproduces not only the filament
elongation, but also the mass of molecular gas detected. At radii $\le
40\,\rm{kpc}$, the mean velocities and the velocity dispersions are
also in good agreement, but at larger radius, the model fails to
reproduce the negative velocities and most of the time it
underestimates the velocity dispersions by a factor of 1.5, 2 and 1.3
for regions r1, r2, and r3 respectively.  The negative velocities can
be understood if the gas is still being uplifted by the rising
bubble. In the simulation however, at temperature below
$10^6\,\rm{K}$, the cooling gas decouples from the ambient gas and
falls towards the centre. This discrepancy would be removed if another
physical process, like viscosity or magnetic fields (Fabian et al.,
2008) increase the coupling between the different gas phases.

\subsection{Line ratios}
Among the 12 regions detected here, only three have been detected in
CO(1--0) and CO(2--1). All the others are detected in CO(2--1) but not
in CO(1--0). One region is detected in CO(1--0) and not in
CO(2--1). As already mentioned, this could be explained by effects of
beam dilution that makes the CO(1--0) emission from a clumpy molecular
gas more difficult to detect than the CO(2--1) emission in
general. Nevertheless, when closer to the centre, the filaments are
widespread and overlap both the 3mm and 1mm beams. However, the
emission seen in each wavelength may not come from the same part of
the filaments. Evidence for that is the different line shapes seen in
region r4, indicating CO(1--0) and CO(2--1) emission from dynamically
and spatially different regions. This also explains the clear CO(1--0)
detection alone in region r19: the region that dominates the emission
lies in the CO(1--0) beam but not in the CO(2--1) one. In conclusion,
the CO(2--1) to (1--0) line ratios are very uncertain, as are the
parameters derived from these values.\\ The three regions with best
signal-to-noise ratios (r4, r15, and r20) were detected in both
CO(1-0) and (2-1).  Averaged over these three regions, the I$_{\rm
CO}$ and T$_{\rm mb}$ (2--1/1--0) ratios are both equal to
1.5$\pm$0.3.  Because interferometer maps (Salome et al 2008b; Lim et
al. 2008) show the CO is extended in the direction of the filaments,
the different beam areas at CO(1-0) and (2-1) mean that the apparent
line ratios should be corrected downward by a factor of 2, so that the
corrected Ico and T$_{\rm mb}$ ratios are close to 0.8, that is, of
the order of unity.\\
Simulations with the line escape probability program RADEX (van der
Tak et al. 2007), indicate that these observed CO(2--1)/CO(1--0) line
ratios of order unity can be obtained for gas kinetic temperatures
ranging from 20 to 500 K, for H$_2$ gas densities $>$10$^{2.5-3}$
cm$^{-3}$ and N$_{\rm CO}$/$\Delta$V $>$10$^{15}$ cm$^{-2}$km/s (right
panel of Fig. 2). So this gives very poor constraints on the gas
properties.\\
With the JCMT, Bridges \& Irwin (1998) observed
CO(3-2) emission from the central 21-arcsec region of NGC~1275 and
marginally detected the $^{13}$CO(2-1) inside a 22$''$ beam. They made
a two-component LVG model to reproduce the 
$^{12}$CO(3--2)/$^{12}$CO(2--1)=1.25$\pm$0.25} and
$^{12}$CO(2--1)/$^{12}$CO(1--0)=0.74$\pm$0.11 line ratios. Their best
model is for a cold component (10K) at a density of 10$^3$cm$^{-3}$
and a hot (170K) component at a density of 3.1\,10$^4$cm$^{-3}$.
Furthermore, new interferometer observations at CO(3-2) (Salome et al,
in preparation) indicate that the CO(3-2)/CO(2-1) ratio in
some central filaments is also of order unity. This
indicates that the CO is certainly optically thick in the central
region and that the low-J CO line brightness and excitation
temperatures are all close to the gas kinetic temperature. \\
It is not possible to definitively rule out an optically thin gas in
the outer filament without CO(3-2) observations there. However, given
that the observed (beam diluted) CO(2-1) brightness temperatures are
about 7 mK, the area filling factor f$_{\rm 2D}$ in
the CO(2-1) beam must be $<$10$^{-3}$.  As a check on this reasoning,
for the three filament regions (r4, r15 and r20), the H$_2$ gas masses
(column 8 of Table 1) estimated from the observed CO line intensities
with a standard Milky Way conversion factor, are about 10$^8$
M$\odot$, within the beam.  The beam diameter at CO(2-1) corresponds
to 4kpc, or 1.3\,10$^{22}$cm, so the equivalent smoothed-out H$_2$
density within a 4-kpc diameter sphere containing 10$^8$ M$\odot$ of
H$_2$ gas would be 0.03 cm$^{-3}$, far too low a gas density to
collisionally excite the CO (the CO(1--0) and
CO(2--1) critical densities being 3\,10$^{3}$cm$^{-3}$ and
10$^{4}$cm$^{-3}$ respectively, for a temperature of 100K).\\
Therefore, to reach the gas density of $>$10$^3$ cm$^{-3}$ needed to
make the low-J CO lines optically thick, and to give line ratios close
to unity, the volume filling factor f$_{\rm 3D}$ 
must be $<$3\,10$^{-5}$.  The corresponding area filling factor within
the beam must be $<$10$^{-3}$, and the linear filling
factor f$_{\rm 1D}$=f$_{\rm 2D}^{1/2}$=f$_{\rm
3D}^{1/3}$, perpendicular to the filaments (or along the line of
sight), must be $<$0.03, relative to the 4-kpc beam size.\\ These
upper limits on the filling factors, derived from the CO conversion
factor and the gas density required for excitation, are thus perfectly
consistent with the filling factor upper limits derived above, from
the observed line temperatures and those predicted by the escape
probability programs, for the observed low-J line ratios.  The linear
filling factor of $<$0.03 implies a CO filament width (or CO clump
size) of $<$120/$\sqrt{N}$ pc, N being the number of
clouds inside the beam, which is also consistent with our size upper
limit of $<$450 pc estimated directly from our interferometric maps
(Salome et al. 2008b).\\ There is therefore a coherent picture in
which to interpret the observed low-J CO line ratios.  The key is the
low filling factor, which explains the observed main-beam brightness
temperatures, the high gas density needed for collisional excitation
in the escape probability predictions, the fact that the low-J line
ratios are of order unity, the fact that the low-J lines are optically
thick, and the interferometer upper limits on the CO clump sizes.

\subsection{Molecular filament formation}
The origin of the huge molecular filaments is unknown. We discuss here
two possibilities. Scenario (1): The molecular gas forms {\itshape in
situ}, far from the galaxy's centre, from uplifted warm gas that
cooled down behind rising bubbles, and eventually falls back (Revaz et
al., 2008).  This scenario must explain how to make the H$_2$, which
is difficult to form without dust (Ferland et al. 2009).  Certainly,
the bubble morphology of the filaments and X-ray cavities show that
this gas must be mixed from the centre of the galaxy up to 100\,kpc
outward, and some dust from the galaxy must have enriched the whole
region.  The presence of dust could augment H$_2$ formation rates even
in the presence of X-rays. But the sputtering due to hot ions in the
X-ray emitting gas can destroys unshielded dust in $\leq$ 1 Myr
(Dwek \& Arendt, 1992), so if there is dust in the filaments, it
must have come from the galaxy's inner regions, rather from the
intracluster medium.  Scenario (2): The molecular gas comes from the
inner part of the galaxy, entrained and dragged out by rising bubbles.
The problem in this scenario is to explain how very small and dense
clouds can be uplifted inside warmer gas.  This is difficult to
achieve, even in a highly viscous outflow and there is no such example
in any observed galaxy yet. More modelling is needed to define the
limits of these different scenarios.  Stronger limits on the amount of
dust in the filaments would help constrain the first scenario, and
modelling the momentum transfer to clumps in a viscous flow (with and
without magnetic viscosity) would help constrain the second scenario.

\subsection{Star Formation}
There is no evidence of strong star formation in NGC~1275
filaments. Canning et al. (2010) have identified, with the Hubble
Space Telescope, only two stellar regions in the south and south-east
filaments (∼22 kpc from the nucleus). The predominantly young stellar
population is concentrated in the H$\alpha$ filaments.  The authors
derived a lower limit of 9$\times$10$^8$M$_\odot$ stars in blue
clusters and a maximum lifetime of $\sim$0.75$\times$10$^8$yr, which
gives a very large SFR of $\sim$20 M$_\odot$/yr. We have compared
the amount of molecular gas with the star formation activity in these
filaments The region of interest is indicated as r15 ('Blue-loop'
filament) in the present paper and fills the CO(1--0) beam size.  We
have found a molecular gas mass of $\sim$2$\times$10$^8$M$_\odot$. So 
if the SFR is as large as $\sim$20 M$_\odot$/yr then, the amount of
gas is about 20\% of the amount of recently formed stars. This means
that most of the gas must have already been transformed into stars
during the filament dynamical time (that roughly corresponds to the
age of the stars).  So if the star formation is still taking place in
r15, the star formation efficiency inside this filament is very
high. Indeed, even if the amount of molecular gas has been
underestimated because of the sub-solar metallicity in the filaments
Z$<$0.6Z$\odot$, (Schmidt et al., 2002), the filaments will still
appear to be locally the place of an enhanced star formation activity
with a star formation efficiency closer to ULIRGs than to normal
spiral galaxies. The gas uplifted by the AGN bubble expansion in the
ICM is pushed and compressed. In these overdense regions, the cooling
is more efficient and molecular gas forms. This reservoir could then
be converted very efficiently into stars in places where large scale
dynamical effects can enhance the star formation (like in spiral arms
of normal galaxies). The interface between the AGN cavities and the
intracluster medium seems to be the place of such processes.

%---------------------------------------------------------------------------
\section{Conclusions}
%---------------------------------------------------------------------------

We have detected CO in a very extended filamentary network around
NGC~1275. Up to about 10$^9$M$_\odot$ of molecular gas is detected
inside thin and elongated filaments at distances between 7.8\,kpc
and 50\,kpc from the galaxy centre. It is possible that the massive
filaments seen in CO are an effect of the AGN feedback.  The large
scale motions in the Northern filament are difficult to explain with
free-fall models. An accurate study of the dynamics of the filament
should take into account the viscosity of the warmer gas (some part of
it being uplifted). This will help lower the model predictions of the
cold gas velocities.  These single dish observations do not resolve
the very thin filaments as observed with the HST (Fabian et al. 2008),
so we certainly also mixed several filaments and averaged their
individual velocities. The CO observed linewidths are large.  This
confirms that the emission very likely comes from substructures
unresolved by the 30\,m telescope, possibly molecular cloud complexes
within the filaments.  Finally we reported a velocity structure in the
emitting gas within 2\,kpc of the AGN that could be a rotation disk.
Whether this is the fate of the molecular gas found in very distant
regions (20-50\,kpc) is still an open question.

%%%%%%%%%%%%%%%%%%%%%%%%%%%%%%%%%%%%%%%%%%%%%%%%%%%%%%%%%%%%%%%%%%%
\begin{acknowledgements}
IRAM is supported by INSU/CNRS (France), MPG (Germany) and IGN
(Spain).  We thank the IRAM 30m-telescope operators for their expert
help with the observing.
\end{acknowledgements}
%%%%%%%%%%%%%%%%%%%%%%%%%%%%%%%%%%%%%%%%%%%%%%%%%%%%%%%%%%%%%%%%%%%

%---------------------------------------------------------------------------

%---------------------------------------------------------------------------
\end{document}